\documentclass[aps,prl,twocolumn,superscriptaddress]{revtex4-2}
\usepackage{amsmath}
\usepackage{mathrsfs}
\usepackage{dsfont}
\usepackage{amssymb}
\usepackage{textcomp}
\usepackage{hyperref}
\usepackage{graphicx}
\usepackage{float}
\usepackage{siunitx}
\usepackage{color}
\usepackage[dvipsnames]{xcolor}
\usepackage{bm}
\usepackage{soul}

\DeclareMathOperator\arctanh{arctanh}
\DeclareMathOperator\arccot{arccot}

\begin{document}
\title{Relativistic quantum mechanics of charged vortex particles accelerated in a uniform electric field}
\author{Qi Meng}
\affiliation{Sino-French Institute of Nuclear Engineering and Technology, Sun Yat-Sen University, Zhuhai 519082, China}
\author{Ziqiang Huang}
\affiliation{Sino-French Institute of Nuclear Engineering and Technology, Sun Yat-Sen University, Zhuhai 519082, China}
\author{Xuan Liu}
\affiliation{Sino-French Institute of Nuclear Engineering and Technology, Sun Yat-Sen University, Zhuhai 519082, China}
\author{Wei Ma}
\affiliation{Sino-French Institute of Nuclear Engineering and Technology, Sun Yat-Sen University, Zhuhai 519082, China}
\author{Zhen Yang}
\affiliation{Sino-French Institute of Nuclear Engineering and Technology, Sun Yat-Sen University, Zhuhai 519082, China}
\author{Liang Lu}
\affiliation{Sino-French Institute of Nuclear Engineering and Technology, Sun Yat-Sen University, Zhuhai 519082, China}
\author{Alexander J. Silenko}
\affiliation{Bogoliubov Laboratory of Theoretical Physics, Joint Institute for Nuclear Research, Dubna 141980, Russia}
\author{Pengming Zhang}
\affiliation{School of Physics and Astronomy, Sun Yat-sen University, Zhuhai 519082, China}
\author{Liping Zou}
\email[Contact author.\\]{zoulp5@mail.sysu.edu.cn}
\affiliation{Sino-French Institute of Nuclear Engineering and Technology, Sun Yat-Sen University, Zhuhai 519082, China}

\begin{abstract}
The relativistic quantum-mechanical description of a charged Laguerre-Gauss beam accelerated in a uniform electric field has been fulfilled. Stationary wave eigenfunctions are rigorously derived. 
The evolution of the beam parameters during acceleration is considered in detail. The practically important effect of extraordinary suppression of transverse spreading of the beam is discovered, carefully analyzed, and properly explained. Our results provide direct evidence that vortex particle beams can be accelerated without destroying their intrinsic vortex properties, paving the way for high-energy vortex beam applications.
\end{abstract}
\maketitle

Vortex (twisted) electrons carrying the orbital angular momentum (OAM) have numerous potential applications in fundamental and applied physics (see Refs. \cite{Bliokh_2017,Lloyd_2017,Larocque_2018,Ivanov_2022,Ivanov_2012,Karlovets_2012,Ivanov_2020,Zou_2023,Zou_2021,Meng_2025} and references therein). However, a potential of application of vortex electrons and other vortex particles significantly depends on their energy ranges. Therefore, a successful acceleration of vortex beams would be a very promising achievement. The simplest acceleration can be performed by a uniform electric field \cite{Karlovets_2021}. For this purpose, an axisymmetric electromagnetic lens can also be used \cite{Baturin_2022}. The injection and acceleration of electrons can also be carried out in a twisted laser beam \cite{Tang_2024,Dyatlov_2024,Ababekri_2024}. Production of high-energy particle beams carrying the OAM is now one of the most actual problems of physics of charged vortex particles~\cite{Li_2024,Wu_2022,Ababekri_2024b,Lu_2025,Zhao_2021,Zou_2024}. However, a needed relativistic quantum-mechanical description of the acceleration of such particles is absent. This description is carried out in the present paper.

We use the system of units $\hbar=1,~c=1$ with $\hbar$ and $c$ explicitly included when needed for clarity. 

To perform a rigorous quantum-mechanical analysis, we use the initial Dirac-Pauli Hamiltonian for a charged particle in a uniform electric field collinear to the $z$ axis:
\begin{equation} \begin{array}{c} {\cal H}=\beta m+{\cal E}+{\cal
O},\qquad {\cal E}=e\Phi, \\ {\cal O}=c\bm{\alpha}\cdot\bm{p}+i\mu'\bm{\gamma}\cdot\bm{ E},\qquad \Phi=\Phi_0-E_z z.\end{array}\label{DireqIII} \end{equation}
Here $m$ is the particle rest mass, $\mu'$ is an anomalous magnetic moment and standard denotations of the Dirac matrices (see Ref. \cite{Berestetskii_1982}) are applied. A particle is accelerated along the $z$ axis when $eE_z=|eE_z|>0$.

Our analysis is applicable not only for an infinite space with a uniform electric field but also covers  the transition of a charged particle beam from the free space or a solenoid with a uniform magnetic field to the electric field region. In the latter case, we use the ``hard-edge'' approximation and suppose that there is the sharp boundary between two media at $z=0$. As follows from QM, the wave function and its first derivative should be continuous on the boundary. In particular, the beam width and its first derivative, $w(0)$ and $w'(0)$, should be continuous.

To obtain a clear Schr\"{o}dinger picture of relativistic quantum mechanics (QM), it is convenient to fulfill the relativistic Foldy-Wouthuysen (FW) transformation \cite{Silenko_2003,Silenko_2008,Silenko_2015,Silenko_2015b,Silenko_2025}. The weak-field approximation is applicable when all terms in the FW Hamiltonian containing commutators of ${\cal E}$ are much less than $mc^2$. When $[{\cal O},{\cal E}]=0$, the FW transformation is exact \cite{Silenko_2003}. Therefore, the term ${\cal E}$ is not covered by the weak-field approximation. When this approximation is applied, the relativistic FW transformation for a particle in a uniform electric field results in \cite{Silenko_2003,Silenko_2025}
\begin{equation}\begin{array}{c} \frac{\partial\psi_{FW}}{\partial t}={\cal H}_{FW}\psi_{FW},\\
{\cal H}_{FW}=\beta\epsilon+e\Phi+\left(\frac{\mu_0m}{\epsilon
   +m}+\mu'\right)\frac{1}{\epsilon}\Bigl[\bm\Sigma\cdot(\bm p
\times\bm E)\Bigr],
\end{array} \label{eq33new} \end{equation}
where $\mu_0=e\hbar/(2m)$ is the Dirac magnetic moment and $\epsilon=\sqrt{m^2+\bm{p}^2}$.

In the present study, we consider stationary states (${\cal H}_{FW}\psi_{FW}=\mathbb{E}_0\psi_{FW}$). Such states can be realized only in stationary external fields. Any  vortex beam is a continuum of coherent partial de Broglie waves with the same total energy. In stationary states, the total energy is constant, and the coherence of the beam is not violated. As a result, stationary Laguerre-Gauss (LG) beams remain coherent in any stationary electric, magnetic, and gravitational fields. We do not consider nonstationary LG beams studied in Refs. \cite{Sizykh_2024,Sizykh_2024b}.

For any partial beam in a uniform electric field, the spin is quantized along the axis normal to the vectors $\bm E$ and $\bm p$. For a vortex beam, the spin quantization is only approximate and only in the case of $L_z\gg\hbar$. In this case, the spin quantization axis is radial ($s_r\approx\pm1/2$). The states with $s_r\approx\pm1/2$ are mixed due to a nonzero radial particle momentum.

As a rule, practically used vortex beams satisfy the paraxial approximation ($|\bm p_\bot|\ll |p_z|$). In this case, a transition to the second-order paraxial equation is rather convenient. We can mention that the transition from the first-order to the second-order relativistic quantum-mechanical equation can be fulfilled with any required precision \cite{Silenko_2001,Silenko_2022,Silenko_2025}. In the considered case, we obtain \cite{Silenko_2025}
\begin{equation}
\begin{array}{c}
\left[\left(i\frac{\partial}{\partial t}-\mathfrak{E}\right)^2-\bm{p}^2-m^2\right]\psi=0,\\
\mathfrak{E}=e\Phi+\left(\frac{\mu_0m}{\epsilon
   +m}+\mu'\right)\frac{1}{\epsilon}\Bigl[\bm\Sigma\cdot(\bm p
\times\bm E)\Bigr].
\end{array}
\label{eqgenef}
\end{equation}

For stationary solutions,
\begin{equation}\begin{array}{c}\label{eqeq_2o}
\left[(\mathbb{E}_0-\mathfrak{E})^2-\mathbf{p}^2-m^2\right]\psi=0,\\ \mathbb{E}_0=\epsilon_0+e\Phi_0=\sqrt{m^2+p_0^2}+e\Phi_0,
\end{array}\end{equation}
where $\mathbb{E}_0$ is the conserved total energy. The spin-dependent term in the formula for $\mathfrak{E}$ is rather small compared with $e\Phi$ and can be neglected ($\mathfrak{E}=e\Phi$).  
In the paraxial approximation, $\bm{p}^2\approx p{p}_z+\bm{p}_{\bot}^2/2$ and Eq. (\ref{eqeq_2o}) takes the form 
\begin{equation}
\left[2(\mathbb{E}_0-e\Phi)^2-(2p{p}_z+\bm{p}_{\bot}^2)-2m^2\right]\psi=0.
\label{eqgenep}\end{equation}
Evidently, 
\begin{equation}
\bm p^2=(\mathbb{E}_0-e\Phi)^2-m^2=p^2_0+2\epsilon_0|eE_z|z+e^2E_z^2z^2.
\label{eqforep}\end{equation} 
To proceed to the paraxial equation, one needs to introduce the wavenumbers $k\equiv k(z)=p/\hbar$ and $k_0=p_0/\hbar$. 
Here 
\begin{equation}
k(z)=k_0\sqrt{1+2K_1z+K_2^2z^2}
\label{eqfonep}\end{equation} and 
\begin{equation}\label{eq:eq_K1K2}
	K_1=\frac{\epsilon_0|eE_z|}{c^2p_0^2},\qquad
	K_2=\frac{|eE_z|}{cp_0}.
\end{equation} $k(z)$ should not be confused with $k_z$. $K_1>K_2$ because $\epsilon_0=\sqrt{m^2+p_0^2}>p_0$. The substitution $\psi=\exp{[i\int{k(z)dz}]}\Psi$ (cf. Refs. \cite{Silenko_2019,Zou_2020}) into the equation
$$
\left(2\bm p^2+2i\hbar p\frac{\partial}{\partial z}-\bm{p}_{\bot}^2\right)\psi=0
$$ results in

\begin{equation}
\left[\bm{\nabla}_{\perp}^2+2i k(z)\frac{\partial}{\partial z}\right]\Psi=0.
\label{par}\end{equation}
Unlike the conventional paraxial wave equation in free space, this equation features the wavenumber $k$ that varies with $z$.
\begin{figure*}
	\centering
	\includegraphics[width=\linewidth]{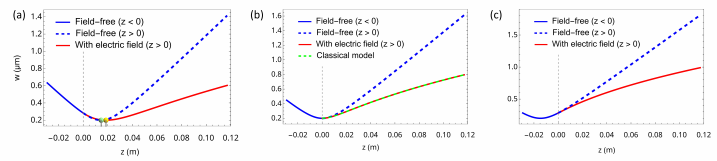}
	\caption{Evolution of the beam width $w(z)$ under a uniform electric field (1.4 MV/m) for initial conditions (a) $w_0' = -2/k_0w_0$, (b) $w_0’=0$, and (c) $w_0' =2/k_0w_0$. The initial kinetic energy is 20 keV and the free-space LG beam waist is $W_0$=200 nm. In (a), the yellow sphere marks the analytical minimum predicted by Eqs. (\ref{eqmin}) and (\ref{zfe}), the green sphere indicates the free-space imaginary focus. The two short gray bars indicate the corresponding minima extracted numerically from the data, which are in good agreement with the analytical prediction.}
	\label{fig:fig_w}
\end{figure*}
\begin{figure*}
	\centering
	\includegraphics[width=\linewidth]{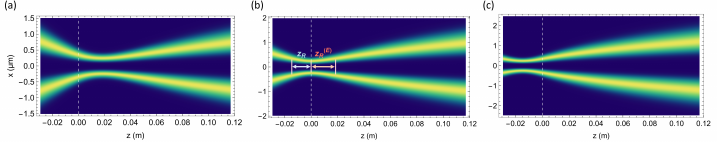}
	\caption{Beam profiles in the $xz$ plane at $y=0$, with parameters identical to those in Fig.~\ref{fig:fig_w}. The example corresponds to $n=0$, $\ell=-3$. In (b), the Rayleigh distance in free space, $z_R$, along with the modified Rayleigh distance in the electric field, $z_R^{(E)}$ (see Eq.~(\ref{zR})), are marked.}
	\label{fig:fig_2}
\end{figure*}

The paraxial equation in free space,
\begin{equation}
\left(\bm{\nabla}_{\perp}^2+2i k\frac{\partial}{\partial z}\right)\Psi=0,
\label{parfr}\end{equation}
($k=const$) has the following solution for LG beams \cite{Barnett_2017,Allen_1992,Plick_2015}:
\begin{equation}
\begin{array}{c}
\Psi=\mathbb{A}\exp{(i\Phi)},\qquad \int{\Psi^\dag\Psi rdrd\phi}=1,\\
\mathbb{A}=\frac{C_{n\ell}}{w(z)}\left(\frac{\sqrt2r}{w(z)}\right)^{|\ell|}
L_n^{|\ell|}\left(\frac{2r^2}{w^2(z)}\right)\exp{\left(-\frac{r^2}{w^2(z)}\right)}\eta,\\
\Phi=l\phi+\frac{kr^2}{2R(z)}-\Phi_G(z),\qquad C_{n\ell}=\sqrt{\frac{2n!}{\pi(n+|\ell|)!}},
\end{array}
\label{eq33new}
\end{equation} where
\begin{equation}
\begin{array}{c}
w(z)=w_0\sqrt{1+\frac{z^2}{z_R^2}},\quad
R(z)=z+\frac{z_R^2}{z},\quad
z_R=\frac{kw_0^2}{2},\\ 
\Phi_G(z)=N\arctan{\left(\frac{z}{z_R}\right)},\quad N=2n+|\ell|+1,
\end{array}
\label{eqaddit}
\end{equation}
the real functions $\mathbb{A}$ and $\Phi$ define the amplitude and phase,
$k$ is the beam wavenumber, $w_0$ is the beam
waist (minimum beam width), $R(z)$ is the radius of curvature of the wavefront, $\Phi_G(z)$ is the Gouy phase, $z_R$ is the Rayleigh diffraction length, $L_n^{|\ell|}$ is the generalized Laguerre polynomial, and $n = 0, 1, 2,\dots$ is the radial quantum number. The spin function $\eta$ is an eigenfunction of the Pauli operator $\sigma_z$: $\sigma_z\eta^\pm=\pm\eta^\pm$, $\eta^+=\left(\begin{array}{c} 1 \\ 0 \end{array}\right)$, $\eta^-=\left(\begin{array}{c} 0 \\ 1 \end{array}\right)$.
$\Psi$ is a spinor and, evidently, is not an eigenfunction of the operator $p_z$. Therefore, the free-space wave function (\ref{eq33new}), (\ref{eqaddit}) characterizes a beam formed by partial waves with
different $p_z$.

The free-space LG beam is perfectly symmetric. When the beam waist is shifted relative to the point $z=0$ and 
$w(0)=w_0$, $w'(0)=w_0'$, the focus position $z_f$ and the beam waist $W_0$ are defined by
\begin{equation}
X=\frac{z_f}{z_R}=\frac{kw_0w_0'}{2},\quad W_0=\frac{w_0}{\sqrt{1+X^2}},\quad z_R=\frac{kW_0^2}{2}.
\end{equation} In the considered case, the quantities $X,\,z_f$, and $w_0'$ are positive and negative for real and imaginary focuses, respectively.

Our derivation shows that the eigenfunction of the LG beam in the uniform electric field has the form (\ref{eq33new}) with replacing $k\rightarrow k_0$ in the phase $\Phi$. However, the beam parameters are not defined by Eq. (\ref{eqaddit}). It is obtained in Sec.~I of the Supplemental Material~\cite{supplement}, that they satisfy the relations
\begin{subequations}
\begin{align}
	&\frac{k_0}{R(z)}=\frac{k(z)w'(z)}{w(z)},\label{eq:eq_fpre}\\
	&\left[\frac{ik_0}{R(z)}-\frac{2}{w(z)^2}\right]^2+ik(z)\left[\frac{ik_0}{R(z)}-\frac{2}{w(z)^2}\right]'=0,\label{eq:eq_f}\\
	&\Phi'_G(z)=\frac{2(2n+|\ell|+1)}{k(z)w(z)^2}.\label{eq:eq_gouy}
\end{align}
\end{subequations}

Setting $f(z)=\frac{ik_0}{R(z)}-\frac{2}{w(z)^2}$ leads  Eq.~(\ref{eq:eq_f}) to the form
\begin{equation}\label{eq:eq_f2}
f(z)^2+ik(z)f'(z)=0
\end{equation}
Imposing the boundary conditions $w(0)=w_0$, $w'(0)=w_0'$, and using Eq.~(\ref{eq:eq_fpre}), we obtain the corresponding expression for $f(0)$:
\[
f(0)=\frac{ik_0w_0'}{w_0}-\frac{2}{w_0^2}.
\]
The solution to Eq.~(\ref{eq:eq_f2}) for $f(z)$ is given by
\begin{equation}\label{eq:sol_f}
f(z)=\frac{k_0K_2(ik_0w_0w_0'-2)}{k_0K_2w_0^2+2(2i+k_0w_0w_0')A(z)},
\end{equation}
where
\begin{equation}\label{eq:arctanh}
A(z)=\arctanh\left[\frac{K_2}{2K_1+K_2^2z}\left(\frac{k(z)}{k_0}-1\right)\right]
\end{equation}
and $k(z)$ is given by Eq.~(\ref{eqfonep}).

For $z>0$, the beam is accelerated and the argument of the $\arctanh$ function in Eq.~(\ref{eq:arctanh}) lies within the interval (0,1), ensuring that $A(z)$ remains real. The beam width and radius of curvature are then determined by separating the real and imaginary parts of $f(z)$ in Eq.~(\ref{eq:sol_f}):
\begin{equation}\label{eq:eq_w}
w(z)=w_0\sqrt{\left(1+\frac{2A(z)w_0'}{K_2w_0}\right)^2+\frac{16A(z)^2}{k_0^2K_2^2w_0^4}},
\end{equation}

\begin{equation}\label{eq:eq_R}
R(z)=\frac{2A(z)}{K_2}+\frac{k_0^2w_0^3\left[2A(z)w_0'+K_2w_0\right]}{k_0^2w_0^2w_0'\left[2A(z)w_0'+K_2w_0\right]+8A(z)}.
\end{equation} As shown in the Sec.~II of the Supplemental Material~\cite{supplement}, Eq. (\ref{eq:eq_w}) can be simplified for $w_0'=0$:
\begin{widetext}\begin{equation}
w(z)=w_0\sqrt{1+\frac{4}{k_0^2w_0^4K_2^2}\ln^2{\left|\frac{K_1+K_2^2z+K_2\sqrt{1+2K_1z+K_2^2z^2}}{K_1+K_2}\right|}}.\label{apf}\end{equation}\end{widetext} 

Using Eqs. (\ref{eq:eq_gouy}) and (\ref{eq:eq_w}), the Gouy phase is given by:
\begin{equation}\label{eq:eq_gouy2}
	\Phi_G(z)=\Phi_G(0)+N\arccot\left[\frac{k_0w_0w_0'}{2}+\frac{k_0K_2w_0^2}{4A(z)}\right].
\end{equation}

The beam width $w(z)$ and its evolution for various initial conditions are shown in Fig.~\ref{fig:fig_w}, with the corresponding beam profiles in the $xz$ plane displayed in Fig.~\ref{fig:fig_2}. In the short-distance limit $|eE_z|z\ll p_0$, corresponding to $	K_1z\ll1$, $K_2z\ll1$, the solution in the presence of the electric field can be reduced to the familiar free-space LG solution, as demonstrated in Sec.~III of the Supplemental Material~\cite{supplement}. Beyond this regime, however, the longitudinal field significantly alters the beam dynamics, extraordinarily suppressing its natural spreading and giving rise to distinct focusing behavior that depends sensitively on the initial conditions.

It is proven in Sec.~IV of the Supplemental Material~\cite{supplement} that the beam undergoes initial focusing for $w_0'<0$ (see Fig.~\ref{fig:fig_w} (b)), reaching a minimum width of
\begin{equation}
w_{\rm min}=\frac{w_0}{\sqrt{1+\frac14k_0^2w_0^2w_0'^2}}.
\label{eqmin}\end{equation}
The corresponding focal position is given by
\begin{equation}\begin{array}{c}
z_{\rm f}=\frac{K_1}{K_2^2}\left[\cosh\left(\frac{k_0^2K_2w_0^3w_0'}{4+k_0^2w_0^2w_0'^2}\right)-1\right]\\-\frac{1}{K_2}\sinh\left(\frac{k_0^2K_2w_0^3w_0'}{4+k_0^2w_0^2w_0'^2}\right).
\label{zfe}\end{array}\end{equation}

Two focuses can exist for $w_0'\neq0$. When $w_0'>0$, the beam has a real focus for its free-space part, can have an imaginary one for an electric-field part, and vice versa for $w_0'<0$.

An analog of the Rayleigh range in the electric field, defined as the propagation length over which the beam's cross-sectional area doubles from the focus, is derived from Eq.~(\ref{eq:eq_w}). Assuming the focus at $z=0$, with $w_0'=0$, and $\pi w(z_{R}^{(E)})^2=2\pi w_0^2$, the Rayleigh range is given by:
\begin{equation}
z_{R}^{(E)}=\frac{2K_1}{K_2^2}\sinh\left(\frac{1}{4}k_0K_2w_0^2\right)^2+\frac{\sinh\left(\frac{1}{2}k_0K_2w_0^2\right)}{K_2}.
\label{zR}\end{equation}
This extended Rayleigh range, as compared to the free-space case, is illustrated in Fig.~\ref{fig:fig_2} (b).

\begin{figure}
	\centering
	\includegraphics[width=0.6\linewidth]{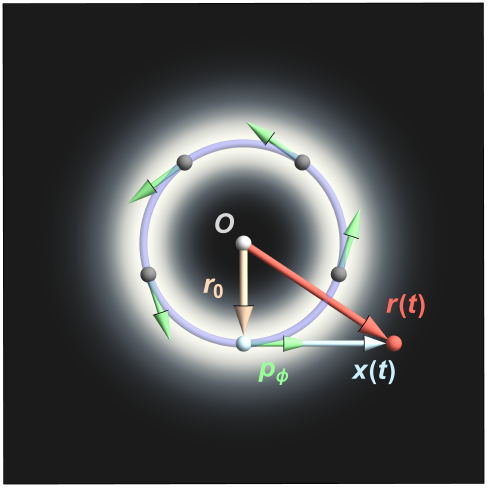}
	\caption{Simple classical model for a LG beam in a uniform electric field.  In the transversal plane, this field does not influence the particle momentum $p_{\phi}$ but decreases the particle velocity. The distance $r(t)$ between the particle and the beam symmetry axis $O$ increases and defines transversal spreading of the beam.}
	\label{fig:fig3}
\end{figure}

Calculated beam behavior can be properly explained in the framework of classical particle physics. A paraxial partial de Broglie wave characterized by transversal momentum $\bm p_\bot$ and total momentum $p\approx p_z$ is equivalent to a particle with the same parameters. At the waist, a LG beam moving in free space can be modeled by a continuum of coherent particle states with identical $p_\phi$ and $p$ and various $\phi$. The probability of particle position azimuth should be uniformly distributed on $\phi$. Let $r_0$ be the distance between the particle and the axis of symmetry of the beam. In the transversal plane, any particle moves freely with constant velocity. At the waist, $v_r=0$ and the direction $x$ of its movement is orthogonal to $\bm r$: $dx=v_\phi dt$. This direction remains unchanged all the time. After time $t$, the azimuth of the particle position changes, the radial velocity becomes nonzero, and the distance to the axis of symmetry increases: $r(t)=\sqrt{r_0^2+v_\phi^2t^2}$ (see Fig. \ref{fig:fig3}). To check the compatibility with wave theory, it is convenient to determine the connection between $r$ and $z$. Since $dz=v_zdt\approx vdt$ and $v_\phi/v=p_\phi/p$, we obtain \begin{equation}
r(z)=\sqrt{r_0^2+\frac{p_\phi^2z^2}{p^2}}.\label{app}\end{equation} Comparison with Eq. (\ref{eqaddit}) shows that the two approaches become equivalent at $r_0=w_0,~p_\phi=2\hbar/w_0$.

If the beam waist is at $z=0$ and the beam is accelerated at $z\geq0$, then $p=\hbar k(z)$ and $p_\phi$ remains unchanged. However, $v_\phi$ decreases in classical and quantum-mechanical pictures because $v_\phi=p_\phi/\epsilon(z)$. In this case, $dx=[p_\phi/\epsilon(z)]dt=[p_\phi/p(z)]dz$. After integration, we obtain \begin{equation}\begin{array}{c}
x=\frac{p_\phi}{p_0}\int_0^z{\frac{dz}{\sqrt{1+2K_1z+K_2^2z^2}}}\\=\frac{p_\phi}{p_0K_2}\ln{\left|\frac{K_1+K_2^2z+K_2\sqrt{1+2K_1z+K_2^2z^2}}{K_1+K_2}\right|}.\label{int}\end{array}\end{equation}
It is amazing that the same substitution as before, $r_0=w_0,~p_\phi=2\hbar/w_0$, results in Eq. (\ref{apf}) for the beam width.

This simple model perfectly explains a fundamental suppression of transverse spreading of the beam in an electric field, but it is approximate. It does not take into account the real quantum-mechanical structure of the LG state. In particular, there is a radial motion even at the waist and the transverse and longitudinal momenta are not definite. Nevertheless, the model leads to the same Eq. (\ref{apf}) which describes an evolution of the beam width.

Vortex beams can enter an electric field not only from free space but also from a solenoid. In this case, $w'_0=0$ for a Landau state, but $w'_0$ can be positive and negative for a LG beam with a spatially oscillating width \cite{Zou_2021,Meng_2025}.

While an electric field extraordinarily suppresses transverse spreading of the beam, the beam focusing can be important. We suppose that it can be performed with magnetic lenses.

In summary, the practically important problem of acceleration of vortex beams is studied. It is shown that charged particle beams keep their coherence and OAMs in a uniform electric field. Stationary LG wave eigenfunctions of relativistic twisted Dirac fermions accelerated in such a field are rigorously derived in the ``hard-edge'' approximation. The evolution of the beam in the field is considered in detail. The important effect of extraordinary suppression of transverse spreading of the beam is discovered, carefully analyzed, and appropriately explained.






\begin{acknowledgments}
The work was supported by the National Key R\&D Program of China No. 2024YFE0109802, the National Natural Science Foundation of China (Grants No.~12175320 and No.~12375084), and Guangdong Basic and Applied Basic Research Foundation (Grant No.~2022A1515010280).
\end{acknowledgments}


\begin{thebibliography}{37}%
	\makeatletter
	\providecommand \@ifxundefined [1]{%
		\@ifx{#1\undefined}
	}%
	\providecommand \@ifnum [1]{%
		\ifnum #1\expandafter \@firstoftwo
		\else \expandafter \@secondoftwo
		\fi
	}%
	\providecommand \@ifx [1]{%
		\ifx #1\expandafter \@firstoftwo
		\else \expandafter \@secondoftwo
		\fi
	}%
	\providecommand \natexlab [1]{#1}%
	\providecommand \enquote  [1]{``#1''}%
	\providecommand \bibnamefont  [1]{#1}%
	\providecommand \bibfnamefont [1]{#1}%
	\providecommand \citenamefont [1]{#1}%
	\providecommand \href@noop [0]{\@secondoftwo}%
	\providecommand \href [0]{\begingroup \@sanitize@url \@href}%
	\providecommand \@href[1]{\@@startlink{#1}\@@href}%
	\providecommand \@@href[1]{\endgroup#1\@@endlink}%
	\providecommand \@sanitize@url [0]{\catcode `\\12\catcode `\$12\catcode
		`\&12\catcode `\#12\catcode `\^12\catcode `\_12\catcode `\%12\relax}%
	\providecommand \@@startlink[1]{}%
	\providecommand \@@endlink[0]{}%
	\providecommand \url  [0]{\begingroup\@sanitize@url \@url }%
	\providecommand \@url [1]{\endgroup\@href {#1}{\urlprefix }}%
	\providecommand \urlprefix  [0]{URL }%
	\providecommand \Eprint [0]{\href }%
	\providecommand \doibase [0]{https://doi.org/}%
	\providecommand \selectlanguage [0]{\@gobble}%
	\providecommand \bibinfo  [0]{\@secondoftwo}%
	\providecommand \bibfield  [0]{\@secondoftwo}%
	\providecommand \translation [1]{[#1]}%
	\providecommand \BibitemOpen [0]{}%
	\providecommand \bibitemStop [0]{}%
	\providecommand \bibitemNoStop [0]{.\EOS\space}%
	\providecommand \EOS [0]{\spacefactor3000\relax}%
	\providecommand \BibitemShut  [1]{\csname bibitem#1\endcsname}%
	\let\auto@bib@innerbib\@empty
	\bibitem [{\citenamefont {Bliokh}\ \emph {et~al.}(2017)\citenamefont {Bliokh},
		\citenamefont {Ivanov}, \citenamefont {Guzzinati}, \citenamefont {Clark},
		\citenamefont {Van~Boxem}, \citenamefont {Béché}, \citenamefont
		{Juchtmans}, \citenamefont {Alonso}, \citenamefont {Schattschneider},
		\citenamefont {Nori},\ and\ \citenamefont {Verbeeck}}]{Bliokh_2017}%
	\BibitemOpen
	\bibfield  {author} {\bibinfo {author} {\bibfnamefont {K.~Y.}\ \bibnamefont
			{Bliokh}}, \bibinfo {author} {\bibfnamefont {I.~P.}\ \bibnamefont {Ivanov}},
		\bibinfo {author} {\bibfnamefont {G.}~\bibnamefont {Guzzinati}}, \bibinfo
		{author} {\bibfnamefont {L.}~\bibnamefont {Clark}}, \bibinfo {author}
		{\bibfnamefont {R.}~\bibnamefont {Van~Boxem}}, \bibinfo {author}
		{\bibfnamefont {A.}~\bibnamefont {Béché}}, \bibinfo {author} {\bibfnamefont
			{R.}~\bibnamefont {Juchtmans}}, \bibinfo {author} {\bibfnamefont {M.~A.}\
			\bibnamefont {Alonso}}, \bibinfo {author} {\bibfnamefont {P.}~\bibnamefont
			{Schattschneider}}, \bibinfo {author} {\bibfnamefont {F.}~\bibnamefont
			{Nori}},\ and\ \bibinfo {author} {\bibfnamefont {J.}~\bibnamefont
			{Verbeeck}},\ }\href {https://doi.org/10.1016/j.physrep.2017.05.006}
	{\bibfield  {journal} {\bibinfo  {journal} {Phys. Rep.}\ }\textbf {\bibinfo
			{volume} {690}},\ \bibinfo {pages} {1} (\bibinfo {year} {2017})}\BibitemShut
	{NoStop}%
	\bibitem [{\citenamefont {Lloyd}\ \emph {et~al.}(2017)\citenamefont {Lloyd},
		\citenamefont {Babiker}, \citenamefont {Thirunavukkarasu},\ and\
		\citenamefont {Yuan}}]{Lloyd_2017}%
	\BibitemOpen
	\bibfield  {author} {\bibinfo {author} {\bibfnamefont {S.~M.}\ \bibnamefont
			{Lloyd}}, \bibinfo {author} {\bibfnamefont {M.}~\bibnamefont {Babiker}},
		\bibinfo {author} {\bibfnamefont {G.}~\bibnamefont {Thirunavukkarasu}},\ and\
		\bibinfo {author} {\bibfnamefont {J.}~\bibnamefont {Yuan}},\ }\href
	{https://doi.org/10.1103/RevModPhys.89.035004} {\bibfield  {journal}
		{\bibinfo  {journal} {Rev. Mod. Phys.}\ }\textbf {\bibinfo {volume} {89}},\
		\bibinfo {pages} {035004} (\bibinfo {year} {2017})}\BibitemShut {NoStop}%
	\bibitem [{\citenamefont {Larocque}\ \emph {et~al.}(2018)\citenamefont
		{Larocque}, \citenamefont {Kaminer}, \citenamefont {Grillo}, \citenamefont
		{Leuchs}, \citenamefont {Padgett}, \citenamefont {Boyd}, \citenamefont
		{Segev},\ and\ \citenamefont {Karimi}}]{Larocque_2018}%
	\BibitemOpen
	\bibfield  {author} {\bibinfo {author} {\bibfnamefont {H.}~\bibnamefont
			{Larocque}}, \bibinfo {author} {\bibfnamefont {I.}~\bibnamefont {Kaminer}},
		\bibinfo {author} {\bibfnamefont {V.}~\bibnamefont {Grillo}}, \bibinfo
		{author} {\bibfnamefont {G.}~\bibnamefont {Leuchs}}, \bibinfo {author}
		{\bibfnamefont {M.~J.}\ \bibnamefont {Padgett}}, \bibinfo {author}
		{\bibfnamefont {R.~W.}\ \bibnamefont {Boyd}}, \bibinfo {author}
		{\bibfnamefont {M.}~\bibnamefont {Segev}},\ and\ \bibinfo {author}
		{\bibfnamefont {E.}~\bibnamefont {Karimi}},\ }\href
	{https://doi.org/10.1080/00107514.2017.1418046} {\bibfield  {journal}
		{\bibinfo  {journal} {Contemp. Phys.}\ }\textbf {\bibinfo {volume} {59}},\
		\bibinfo {pages} {126} (\bibinfo {year} {2018})}\BibitemShut {NoStop}%
	\bibitem [{\citenamefont {Ivanov}(2022)}]{Ivanov_2022}%
	\BibitemOpen
	\bibfield  {author} {\bibinfo {author} {\bibfnamefont {I.~P.}\ \bibnamefont
			{Ivanov}},\ }\href {https://doi.org/doi.org/10.1016/j.ppnp.2022.103987}
	{\bibfield  {journal} {\bibinfo  {journal} {Prog. Part. Nucl. Phys.}\
		}\textbf {\bibinfo {volume} {127}},\ \bibinfo {pages} {103987} (\bibinfo
		{year} {2022})}\BibitemShut {NoStop}%
	\bibitem [{\citenamefont {Ivanov}(2012)}]{Ivanov_2012}%
	\BibitemOpen
	\bibfield  {author} {\bibinfo {author} {\bibfnamefont {I.~P.}\ \bibnamefont
			{Ivanov}},\ }\href {https://doi.org/10.1103/PhysRevD.85.076001} {\bibfield
		{journal} {\bibinfo  {journal} {Phys. Rev. D}\ }\textbf {\bibinfo {volume}
			{85}},\ \bibinfo {pages} {076001} (\bibinfo {year} {2012})}\BibitemShut
	{NoStop}%
	\bibitem [{\citenamefont {Karlovets}(2012)}]{Karlovets_2012}%
	\BibitemOpen
	\bibfield  {author} {\bibinfo {author} {\bibfnamefont {D.~V.}\ \bibnamefont
			{Karlovets}},\ }\href {https://doi.org/10.1103/PhysRevA.86.062102} {\bibfield
		{journal} {\bibinfo  {journal} {Phys. Rev. A}\ }\textbf {\bibinfo {volume}
			{86}},\ \bibinfo {pages} {062102} (\bibinfo {year} {2012})}\BibitemShut
	{NoStop}%
	\bibitem [{\citenamefont {Ivanov}\ \emph {et~al.}(2020)\citenamefont {Ivanov},
		\citenamefont {Korchagin}, \citenamefont {Pimikov},\ and\ \citenamefont
		{Zhang}}]{Ivanov_2020}%
	\BibitemOpen
	\bibfield  {author} {\bibinfo {author} {\bibfnamefont {I.~P.}\ \bibnamefont
			{Ivanov}}, \bibinfo {author} {\bibfnamefont {N.}~\bibnamefont {Korchagin}},
		\bibinfo {author} {\bibfnamefont {A.}~\bibnamefont {Pimikov}},\ and\ \bibinfo
		{author} {\bibfnamefont {P.}~\bibnamefont {Zhang}},\ }\href
	{https://doi.org/10.1103/PhysRevLett.124.192001} {\bibfield  {journal}
		{\bibinfo  {journal} {Phys. Rev. Lett.}\ }\textbf {\bibinfo {volume} {124}},\
		\bibinfo {pages} {192001} (\bibinfo {year} {2020})}\BibitemShut {NoStop}%
	\bibitem [{\citenamefont {Zou}\ \emph {et~al.}(2023)\citenamefont {Zou},
		\citenamefont {Zhang}, \citenamefont {Silenko},\ and\ \citenamefont
		{Lu}}]{Zou_2023}%
	\BibitemOpen
	\bibfield  {author} {\bibinfo {author} {\bibfnamefont {L.}~\bibnamefont
			{Zou}}, \bibinfo {author} {\bibfnamefont {P.}~\bibnamefont {Zhang}}, \bibinfo
		{author} {\bibfnamefont {A.~J.}\ \bibnamefont {Silenko}},\ and\ \bibinfo
		{author} {\bibfnamefont {L.}~\bibnamefont {Lu}},\ }\href
	{https://doi.org/doi.org/10.1016/j.xinn.2023.100432} {\bibfield  {journal}
		{\bibinfo  {journal} {Innovation}\ }\textbf {\bibinfo {volume} {4}},\
		\bibinfo {pages} {100432} (\bibinfo {year} {2023})}\BibitemShut {NoStop}%
	\bibitem [{\citenamefont {Zou}\ \emph {et~al.}(2021)\citenamefont {Zou},
		\citenamefont {Zhang},\ and\ \citenamefont {Silenko}}]{Zou_2021}%
	\BibitemOpen
	\bibfield  {author} {\bibinfo {author} {\bibfnamefont {L.}~\bibnamefont
			{Zou}}, \bibinfo {author} {\bibfnamefont {P.}~\bibnamefont {Zhang}},\ and\
		\bibinfo {author} {\bibfnamefont {A.~J.}\ \bibnamefont {Silenko}},\ }\href
	{https://doi.org/10.1103/PhysRevA.103.L010201} {\bibfield  {journal}
		{\bibinfo  {journal} {Phys. Rev. A}\ }\textbf {\bibinfo {volume} {103}},\
		\bibinfo {pages} {L010201} (\bibinfo {year} {2021})}\BibitemShut {NoStop}%
	\bibitem [{\citenamefont {Meng}\ \emph {et~al.}(2025)\citenamefont {Meng},
		\citenamefont {Liu}, \citenamefont {Ma}, \citenamefont {Yang}, \citenamefont
		{Lu}, \citenamefont {Silenko}, \citenamefont {Zhang},\ and\ \citenamefont
		{Zou}}]{Meng_2025}%
	\BibitemOpen
	\bibfield  {author} {\bibinfo {author} {\bibfnamefont {Q.}~\bibnamefont
			{Meng}}, \bibinfo {author} {\bibfnamefont {X.}~\bibnamefont {Liu}}, \bibinfo
		{author} {\bibfnamefont {W.}~\bibnamefont {Ma}}, \bibinfo {author}
		{\bibfnamefont {Z.}~\bibnamefont {Yang}}, \bibinfo {author} {\bibfnamefont
			{L.}~\bibnamefont {Lu}}, \bibinfo {author} {\bibfnamefont {A.~J.}\
			\bibnamefont {Silenko}}, \bibinfo {author} {\bibfnamefont {P.}~\bibnamefont
			{Zhang}},\ and\ \bibinfo {author} {\bibfnamefont {L.}~\bibnamefont {Zou}},\
	}\href {https://doi.org/10.1103/8jrx-p9rz} {\bibfield  {journal} {\bibinfo
			{journal} {Phys. Rev. Res.}\ ,\ } (\bibinfo {year} {2025})},\ \bibinfo {note}
	{accepted for publication, in press}\BibitemShut {NoStop}%
	\bibitem [{\citenamefont {Karlovets}(2021)}]{Karlovets_2021}%
	\BibitemOpen
	\bibfield  {author} {\bibinfo {author} {\bibfnamefont {D.}~\bibnamefont
			{Karlovets}},\ }\href {https://doi.org/10.1088/1367-2630/abeacc} {\bibfield
		{journal} {\bibinfo  {journal} {New J. Phys.}\ }\textbf {\bibinfo {volume}
			{23}},\ \bibinfo {pages} {033048} (\bibinfo {year} {2021})}\BibitemShut
	{NoStop}%
	\bibitem [{\citenamefont {Baturin}\ \emph {et~al.}(2022)\citenamefont
		{Baturin}, \citenamefont {Grosman}, \citenamefont {Sizykh},\ and\
		\citenamefont {Karlovets}}]{Baturin_2022}%
	\BibitemOpen
	\bibfield  {author} {\bibinfo {author} {\bibfnamefont {S.~S.}\ \bibnamefont
			{Baturin}}, \bibinfo {author} {\bibfnamefont {D.~V.}\ \bibnamefont
			{Grosman}}, \bibinfo {author} {\bibfnamefont {G.~K.}\ \bibnamefont
			{Sizykh}},\ and\ \bibinfo {author} {\bibfnamefont {D.~V.}\ \bibnamefont
			{Karlovets}},\ }\href {https://doi.org/10.1103/PhysRevA.106.042211}
	{\bibfield  {journal} {\bibinfo  {journal} {Phys. Rev. A}\ }\textbf {\bibinfo
			{volume} {106}},\ \bibinfo {pages} {042211} (\bibinfo {year}
		{2022})}\BibitemShut {NoStop}%
	\bibitem [{\citenamefont {Tang}\ \emph {et~al.}(2024)\citenamefont {Tang},
		\citenamefont {Hao},\ and\ \citenamefont {Shi}}]{Tang_2024}%
	\BibitemOpen
	\bibfield  {author} {\bibinfo {author} {\bibfnamefont {X.}~\bibnamefont
			{Tang}}, \bibinfo {author} {\bibfnamefont {J.}~\bibnamefont {Hao}},\ and\
		\bibinfo {author} {\bibfnamefont {Y.}~\bibnamefont {Shi}},\ }\href
	{https://doi.org/10.1017/hpl.2024.56} {\bibfield  {journal} {\bibinfo
			{journal} {High Power Laser Sci. Eng.}\ }\textbf {\bibinfo {volume} {12}},\
		\bibinfo {pages} {e86} (\bibinfo {year} {2024})}\BibitemShut {NoStop}%
	\bibitem [{\citenamefont {Dyatlov}\ \emph {et~al.}(2024)\citenamefont
		{Dyatlov}, \citenamefont {Bleko}, \citenamefont {Cherepanov}, \citenamefont
		{Kobets}, \citenamefont {Martyanov}, \citenamefont {Nozdrin}, \citenamefont
		{Sergeev}, \citenamefont {Sheremet}, \citenamefont {Zhemchugov},\ and\
		\citenamefont {Karlovets}}]{Dyatlov_2024}%
	\BibitemOpen
	\bibfield  {author} {\bibinfo {author} {\bibfnamefont {A.~S.}\ \bibnamefont
			{Dyatlov}}, \bibinfo {author} {\bibfnamefont {V.~V.}\ \bibnamefont {Bleko}},
		\bibinfo {author} {\bibfnamefont {K.~V.}\ \bibnamefont {Cherepanov}},
		\bibinfo {author} {\bibfnamefont {V.~V.}\ \bibnamefont {Kobets}}, \bibinfo
		{author} {\bibfnamefont {M.~A.}\ \bibnamefont {Martyanov}}, \bibinfo {author}
		{\bibfnamefont {M.~A.}\ \bibnamefont {Nozdrin}}, \bibinfo {author}
		{\bibfnamefont {A.~N.}\ \bibnamefont {Sergeev}}, \bibinfo {author}
		{\bibfnamefont {N.~E.}\ \bibnamefont {Sheremet}}, \bibinfo {author}
		{\bibfnamefont {A.~S.}\ \bibnamefont {Zhemchugov}},\ and\ \bibinfo {author}
		{\bibfnamefont {D.~V.}\ \bibnamefont {Karlovets}},\ }in\ \href
	{https://doi.org/10.1109/ICLO59702.2024.10624179} {\emph {\bibinfo
			{booktitle} {2024 International Conference Laser Optics (ICLO)}}}\ (\bibinfo
	{publisher} {IEEE},\ \bibinfo {address} {Saint Petersburg},\ \bibinfo {year}
	{2024})\ pp.\ \bibinfo {pages} {438--438}\BibitemShut {NoStop}%
	\bibitem [{\citenamefont {Ababekri}\ \emph
		{et~al.}(2024{\natexlab{a}})\citenamefont {Ababekri}, \citenamefont {Wang},
		\citenamefont {Guo}, \citenamefont {Li},\ and\ \citenamefont
		{Li}}]{Ababekri_2024}%
	\BibitemOpen
	\bibfield  {author} {\bibinfo {author} {\bibfnamefont {M.}~\bibnamefont
			{Ababekri}}, \bibinfo {author} {\bibfnamefont {Y.}~\bibnamefont {Wang}},
		\bibinfo {author} {\bibfnamefont {R.-T.}\ \bibnamefont {Guo}}, \bibinfo
		{author} {\bibfnamefont {Z.-P.}\ \bibnamefont {Li}},\ and\ \bibinfo {author}
		{\bibfnamefont {J.-X.}\ \bibnamefont {Li}},\ }\href
	{https://doi.org/10.1103/PhysRevA.110.052207} {\bibfield  {journal} {\bibinfo
			{journal} {Phys. Rev. A}\ }\textbf {\bibinfo {volume} {110}},\ \bibinfo
		{pages} {052207} (\bibinfo {year} {2024}{\natexlab{a}})}\BibitemShut
	{NoStop}%
	\bibitem [{\citenamefont {Li}\ \emph {et~al.}(2024)\citenamefont {Li},
		\citenamefont {Liu}, \citenamefont {Liu}, \citenamefont {Ji},\ and\
		\citenamefont {Ivanov}}]{Li_2024}%
	\BibitemOpen
	\bibfield  {author} {\bibinfo {author} {\bibfnamefont {Z.}~\bibnamefont
			{Li}}, \bibinfo {author} {\bibfnamefont {S.}~\bibnamefont {Liu}}, \bibinfo
		{author} {\bibfnamefont {B.}~\bibnamefont {Liu}}, \bibinfo {author}
		{\bibfnamefont {L.}~\bibnamefont {Ji}},\ and\ \bibinfo {author}
		{\bibfnamefont {I.~P.}\ \bibnamefont {Ivanov}},\ }\href
	{https://doi.org/10.1103/PhysRevLett.133.265001} {\bibfield  {journal}
		{\bibinfo  {journal} {Phys. Rev. Lett.}\ }\textbf {\bibinfo {volume} {133}},\
		\bibinfo {pages} {265001} (\bibinfo {year} {2024})}\BibitemShut {NoStop}%
	\bibitem [{\citenamefont {Wu}\ \emph {et~al.}(2022)\citenamefont {Wu},
		\citenamefont {Gargiulo}, \citenamefont {Carbone}, \citenamefont {Keitel},\
		and\ \citenamefont {P\'alffy}}]{Wu_2022}%
	\BibitemOpen
	\bibfield  {author} {\bibinfo {author} {\bibfnamefont {Y.}~\bibnamefont
			{Wu}}, \bibinfo {author} {\bibfnamefont {S.}~\bibnamefont {Gargiulo}},
		\bibinfo {author} {\bibfnamefont {F.}~\bibnamefont {Carbone}}, \bibinfo
		{author} {\bibfnamefont {C.~H.}\ \bibnamefont {Keitel}},\ and\ \bibinfo
		{author} {\bibfnamefont {A.}~\bibnamefont {P\'alffy}},\ }\href
	{https://doi.org/10.1103/PhysRevLett.128.162501} {\bibfield  {journal}
		{\bibinfo  {journal} {Phys. Rev. Lett.}\ }\textbf {\bibinfo {volume} {128}},\
		\bibinfo {pages} {162501} (\bibinfo {year} {2022})}\BibitemShut {NoStop}%
	\bibitem [{\citenamefont {Ababekri}\ \emph
		{et~al.}(2024{\natexlab{b}})\citenamefont {Ababekri}, \citenamefont {Zhou},
		\citenamefont {Guo}, \citenamefont {Ren}, \citenamefont {Kou}, \citenamefont
		{Zhao}, \citenamefont {Li},\ and\ \citenamefont {Li}}]{Ababekri_2024b}%
	\BibitemOpen
	\bibfield  {author} {\bibinfo {author} {\bibfnamefont {M.}~\bibnamefont
			{Ababekri}}, \bibinfo {author} {\bibfnamefont {J.-L.}\ \bibnamefont {Zhou}},
		\bibinfo {author} {\bibfnamefont {R.-T.}\ \bibnamefont {Guo}}, \bibinfo
		{author} {\bibfnamefont {Y.-Z.}\ \bibnamefont {Ren}}, \bibinfo {author}
		{\bibfnamefont {Y.-H.}\ \bibnamefont {Kou}}, \bibinfo {author} {\bibfnamefont
			{Q.}~\bibnamefont {Zhao}}, \bibinfo {author} {\bibfnamefont {Z.-P.}\
			\bibnamefont {Li}},\ and\ \bibinfo {author} {\bibfnamefont {J.-X.}\
			\bibnamefont {Li}},\ }\href {https://doi.org/10.1103/PhysRevD.110.076024}
	{\bibfield  {journal} {\bibinfo  {journal} {Phys. Rev. D}\ }\textbf {\bibinfo
			{volume} {110}},\ \bibinfo {pages} {076024} (\bibinfo {year}
		{2024}{\natexlab{b}})}\BibitemShut {NoStop}%
	\bibitem [{\citenamefont {Lu}\ \emph {et~al.}(2025)\citenamefont {Lu},
		\citenamefont {Guo}, \citenamefont {Ababekri}, \citenamefont {Zhang},
		\citenamefont {Weng}, \citenamefont {Wu}, \citenamefont {Niu},\ and\
		\citenamefont {Li}}]{Lu_2025}%
	\BibitemOpen
	\bibfield  {author} {\bibinfo {author} {\bibfnamefont {Z.-W.}\ \bibnamefont
			{Lu}}, \bibinfo {author} {\bibfnamefont {L.}~\bibnamefont {Guo}}, \bibinfo
		{author} {\bibfnamefont {M.}~\bibnamefont {Ababekri}}, \bibinfo {author}
		{\bibfnamefont {J.-L.}\ \bibnamefont {Zhang}}, \bibinfo {author}
		{\bibfnamefont {X.-F.}\ \bibnamefont {Weng}}, \bibinfo {author}
		{\bibfnamefont {Y.}~\bibnamefont {Wu}}, \bibinfo {author} {\bibfnamefont
			{Y.-F.}\ \bibnamefont {Niu}},\ and\ \bibinfo {author} {\bibfnamefont {J.-X.}\
			\bibnamefont {Li}},\ }\href {https://doi.org/10.1103/PhysRevLett.134.052501}
	{\bibfield  {journal} {\bibinfo  {journal} {Phys. Rev. Lett.}\ }\textbf
		{\bibinfo {volume} {134}},\ \bibinfo {pages} {052501} (\bibinfo {year}
		{2025})}\BibitemShut {NoStop}%
	\bibitem [{\citenamefont {Zhao}\ \emph {et~al.}(2021)\citenamefont {Zhao},
		\citenamefont {Ivanov},\ and\ \citenamefont {Zhang}}]{Zhao_2021}%
	\BibitemOpen
	\bibfield  {author} {\bibinfo {author} {\bibfnamefont {P.}~\bibnamefont
			{Zhao}}, \bibinfo {author} {\bibfnamefont {I.~P.}\ \bibnamefont {Ivanov}},\
		and\ \bibinfo {author} {\bibfnamefont {P.}~\bibnamefont {Zhang}},\ }\href
	{https://doi.org/10.1103/PhysRevD.104.036003} {\bibfield  {journal} {\bibinfo
			{journal} {Phys. Rev. D}\ }\textbf {\bibinfo {volume} {104}},\ \bibinfo
		{pages} {036003} (\bibinfo {year} {2021})}\BibitemShut {NoStop}%
	\bibitem [{\citenamefont {Zou}\ \emph {et~al.}(2024)\citenamefont {Zou},
		\citenamefont {Zhang},\ and\ \citenamefont {Silenko}}]{Zou_2024}%
	\BibitemOpen
	\bibfield  {author} {\bibinfo {author} {\bibfnamefont {L.}~\bibnamefont
			{Zou}}, \bibinfo {author} {\bibfnamefont {P.}~\bibnamefont {Zhang}},\ and\
		\bibinfo {author} {\bibfnamefont {A.~J.}\ \bibnamefont {Silenko}},\ }\href
	{https://doi.org/10.1088/1361-6455/ad23f7} {\bibfield  {journal} {\bibinfo
			{journal} {J. Phys. B}\ }\textbf {\bibinfo {volume} {57}},\ \bibinfo {pages}
		{045401} (\bibinfo {year} {2024})}\BibitemShut {NoStop}%
	\bibitem [{\citenamefont {Berestetskii}\ \emph {et~al.}(1982)\citenamefont
		{Berestetskii}, \citenamefont {Lifshitz},\ and\ \citenamefont
		{Pitaevskii}}]{Berestetskii_1982}%
	\BibitemOpen
	\bibfield  {author} {\bibinfo {author} {\bibfnamefont {V.~B.}\ \bibnamefont
			{Berestetskii}}, \bibinfo {author} {\bibfnamefont {E.~M.}\ \bibnamefont
			{Lifshitz}},\ and\ \bibinfo {author} {\bibfnamefont {L.~P.}\ \bibnamefont
			{Pitaevskii}},\ }\href@noop {} {\emph {\bibinfo {title} {Quantum
				Electrodynamics}}},\ \bibinfo {edition} {2nd}\ ed.\ (\bibinfo  {publisher}
	{Pergamon},\ \bibinfo {address} {Oxford},\ \bibinfo {year}
	{1982})\BibitemShut {NoStop}%
	\bibitem [{\citenamefont {Silenko}(2003)}]{Silenko_2003}%
	\BibitemOpen
	\bibfield  {author} {\bibinfo {author} {\bibfnamefont {A.~J.}\ \bibnamefont
			{Silenko}},\ }\href {https://doi.org/10.1063/1.1579991} {\bibfield  {journal}
		{\bibinfo  {journal} {J. Math. Phys.}\ }\textbf {\bibinfo {volume} {44}},\
		\bibinfo {pages} {2952} (\bibinfo {year} {2003})}\BibitemShut {NoStop}%
	\bibitem [{\citenamefont {Silenko}(2008)}]{Silenko_2008}%
	\BibitemOpen
	\bibfield  {author} {\bibinfo {author} {\bibfnamefont {A.~J.}\ \bibnamefont
			{Silenko}},\ }\href {https://doi.org/10.1103/PhysRevA.77.012116} {\bibfield
		{journal} {\bibinfo  {journal} {Phys. Rev. A}\ }\textbf {\bibinfo {volume}
			{77}},\ \bibinfo {pages} {012116} (\bibinfo {year} {2008})}\BibitemShut
	{NoStop}%
	\bibitem [{\citenamefont {Silenko}(2015{\natexlab{a}})}]{Silenko_2015}%
	\BibitemOpen
	\bibfield  {author} {\bibinfo {author} {\bibfnamefont {A.~J.}\ \bibnamefont
			{Silenko}},\ }\href {https://doi.org/10.1103/PhysRevA.91.012111} {\bibfield
		{journal} {\bibinfo  {journal} {Phys. Rev. A}\ }\textbf {\bibinfo {volume}
			{91}},\ \bibinfo {pages} {012111} (\bibinfo {year}
		{2015}{\natexlab{a}})}\BibitemShut {NoStop}%
	\bibitem [{\citenamefont {Silenko}(2015{\natexlab{b}})}]{Silenko_2015b}%
	\BibitemOpen
	\bibfield  {author} {\bibinfo {author} {\bibfnamefont {A.~J.}\ \bibnamefont
			{Silenko}},\ }\href {https://doi.org/10.1103/PhysRevA.91.022103} {\bibfield
		{journal} {\bibinfo  {journal} {Phys. Rev. A}\ }\textbf {\bibinfo {volume}
			{91}},\ \bibinfo {pages} {022103} (\bibinfo {year}
		{2015}{\natexlab{b}})}\BibitemShut {NoStop}%
	\bibitem [{\citenamefont {Silenko}(2025)}]{Silenko_2025}%
	\BibitemOpen
	\bibfield  {author} {\bibinfo {author} {\bibfnamefont {A.~J.}\ \bibnamefont
			{Silenko}},\ }\href {https://doi.org/10.1103/PhysRevA.111.032210} {\bibfield
		{journal} {\bibinfo  {journal} {Phys. Rev. A}\ }\textbf {\bibinfo {volume}
			{111}},\ \bibinfo {pages} {032210} (\bibinfo {year} {2025})}\BibitemShut
	{NoStop}%
	\bibitem [{\citenamefont {Sizykh}\ \emph
		{et~al.}(2024{\natexlab{a}})\citenamefont {Sizykh}, \citenamefont
		{Chaikovskaia}, \citenamefont {Grosman}, \citenamefont {Pavlov},\ and\
		\citenamefont {Karlovets}}]{Sizykh_2024}%
	\BibitemOpen
	\bibfield  {author} {\bibinfo {author} {\bibfnamefont {G.~K.}\ \bibnamefont
			{Sizykh}}, \bibinfo {author} {\bibfnamefont {A.~D.}\ \bibnamefont
			{Chaikovskaia}}, \bibinfo {author} {\bibfnamefont {D.~V.}\ \bibnamefont
			{Grosman}}, \bibinfo {author} {\bibfnamefont {I.~I.}\ \bibnamefont
			{Pavlov}},\ and\ \bibinfo {author} {\bibfnamefont {D.~V.}\ \bibnamefont
			{Karlovets}},\ }\href {https://doi.org/10.1093/ptep/ptae052} {\bibfield
		{journal} {\bibinfo  {journal} {Prog. Theor. Exp. Phys.}\ }\textbf {\bibinfo
			{volume} {2024}},\ \bibinfo {pages} {053A02} (\bibinfo {year}
		{2024}{\natexlab{a}})}\BibitemShut {NoStop}%
	\bibitem [{\citenamefont {Sizykh}\ \emph
		{et~al.}(2024{\natexlab{b}})\citenamefont {Sizykh}, \citenamefont
		{Chaikovskaia}, \citenamefont {Grosman}, \citenamefont {Pavlov},\ and\
		\citenamefont {Karlovets}}]{Sizykh_2024b}%
	\BibitemOpen
	\bibfield  {author} {\bibinfo {author} {\bibfnamefont {G.~K.}\ \bibnamefont
			{Sizykh}}, \bibinfo {author} {\bibfnamefont {A.~D.}\ \bibnamefont
			{Chaikovskaia}}, \bibinfo {author} {\bibfnamefont {D.~V.}\ \bibnamefont
			{Grosman}}, \bibinfo {author} {\bibfnamefont {I.~I.}\ \bibnamefont
			{Pavlov}},\ and\ \bibinfo {author} {\bibfnamefont {D.~V.}\ \bibnamefont
			{Karlovets}},\ }\href {https://doi.org/10.1103/PhysRevA.109.L040201}
	{\bibfield  {journal} {\bibinfo  {journal} {Phys. Rev. A}\ }\textbf {\bibinfo
			{volume} {109}},\ \bibinfo {pages} {L040201} (\bibinfo {year}
		{2024}{\natexlab{b}})}\BibitemShut {NoStop}%
	\bibitem [{\citenamefont {Silenko}(2001)}]{Silenko_2001}%
	\BibitemOpen
	\bibfield  {author} {\bibinfo {author} {\bibfnamefont {A.~J.}\ \bibnamefont
			{Silenko}},\ }\href {https://doi.org/10.1134/1.1378894} {\bibfield  {journal}
		{\bibinfo  {journal} {Phys. At. Nucl.}\ }\textbf {\bibinfo {volume} {64}},\
		\bibinfo {pages} {977} (\bibinfo {year} {2001})}\BibitemShut {NoStop}%
	\bibitem [{\citenamefont {Silenko}(2022)}]{Silenko_2022}%
	\BibitemOpen
	\bibfield  {author} {\bibinfo {author} {\bibfnamefont {A.~J.}\ \bibnamefont
			{Silenko}},\ }\href {https://doi.org/10.1142/S0217732322500973} {\bibfield
		{journal} {\bibinfo  {journal} {Mod. Phys. Lett. A}\ }\textbf {\bibinfo
			{volume} {37}},\ \bibinfo {pages} {2250097} (\bibinfo {year}
		{2022})}\BibitemShut {NoStop}%
	\bibitem [{\citenamefont {Silenko}\ \emph {et~al.}(2019)\citenamefont
		{Silenko}, \citenamefont {Zhang},\ and\ \citenamefont {Zou}}]{Silenko_2019}%
	\BibitemOpen
	\bibfield  {author} {\bibinfo {author} {\bibfnamefont {A.~J.}\ \bibnamefont
			{Silenko}}, \bibinfo {author} {\bibfnamefont {P.}~\bibnamefont {Zhang}},\
		and\ \bibinfo {author} {\bibfnamefont {L.}~\bibnamefont {Zou}},\ }\href
	{https://doi.org/10.1103/PhysRevA.100.030101} {\bibfield  {journal} {\bibinfo
			{journal} {Phys. Rev. A}\ }\textbf {\bibinfo {volume} {100}},\ \bibinfo
		{pages} {030101} (\bibinfo {year} {2019})}\BibitemShut {NoStop}%
	\bibitem [{\citenamefont {Zou}\ \emph {et~al.}(2020)\citenamefont {Zou},
		\citenamefont {Zhang},\ and\ \citenamefont {Silenko}}]{Zou_2020}%
	\BibitemOpen
	\bibfield  {author} {\bibinfo {author} {\bibfnamefont {L.}~\bibnamefont
			{Zou}}, \bibinfo {author} {\bibfnamefont {P.}~\bibnamefont {Zhang}},\ and\
		\bibinfo {author} {\bibfnamefont {A.~J.}\ \bibnamefont {Silenko}},\ }\href
	{https://doi.org/10.1088/1361-6471/ab7a88} {\bibfield  {journal} {\bibinfo
			{journal} {J. Phys. G: Nucl. Part. Phys.}\ }\textbf {\bibinfo {volume}
			{47}},\ \bibinfo {pages} {055003} (\bibinfo {year} {2020})}\BibitemShut
	{NoStop}%
	\bibitem [{\citenamefont {Barnett}\ \emph {et~al.}(2017)\citenamefont
		{Barnett}, \citenamefont {Babiker},\ and\ \citenamefont
		{Padgett}}]{Barnett_2017}%
	\BibitemOpen
	\bibfield  {author} {\bibinfo {author} {\bibfnamefont {S.~M.}\ \bibnamefont
			{Barnett}}, \bibinfo {author} {\bibfnamefont {M.}~\bibnamefont {Babiker}},\
		and\ \bibinfo {author} {\bibfnamefont {M.~J.}\ \bibnamefont {Padgett}},\
	}\href {https://doi.org/10.1098/rsta.2015.0444} {\bibfield  {journal}
		{\bibinfo  {journal} {Phil. Trans. R. Soc. A}\ }\textbf {\bibinfo {volume}
			{375}},\ \bibinfo {pages} {20150444} (\bibinfo {year} {2017})}\BibitemShut
	{NoStop}%
	\bibitem [{\citenamefont {Allen}\ \emph {et~al.}(1992)\citenamefont {Allen},
		\citenamefont {Beijersbergen}, \citenamefont {Spreeuw},\ and\ \citenamefont
		{Woerdman}}]{Allen_1992}%
	\BibitemOpen
	\bibfield  {author} {\bibinfo {author} {\bibfnamefont {L.}~\bibnamefont
			{Allen}}, \bibinfo {author} {\bibfnamefont {M.~W.}\ \bibnamefont
			{Beijersbergen}}, \bibinfo {author} {\bibfnamefont {R.~J.~C.}\ \bibnamefont
			{Spreeuw}},\ and\ \bibinfo {author} {\bibfnamefont {J.~P.}\ \bibnamefont
			{Woerdman}},\ }\href {https://doi.org/10.1103/PhysRevA.45.8185} {\bibfield
		{journal} {\bibinfo  {journal} {Phys. Rev. A}\ }\textbf {\bibinfo {volume}
			{45}},\ \bibinfo {pages} {8185} (\bibinfo {year} {1992})}\BibitemShut
	{NoStop}%
	\bibitem [{\citenamefont {Plick}\ and\ \citenamefont
		{Krenn}(2015)}]{Plick_2015}%
	\BibitemOpen
	\bibfield  {author} {\bibinfo {author} {\bibfnamefont {W.~N.}\ \bibnamefont
			{Plick}}\ and\ \bibinfo {author} {\bibfnamefont {M.}~\bibnamefont {Krenn}},\
	}\href {https://doi.org/10.1103/PhysRevA.92.063841} {\bibfield  {journal}
		{\bibinfo  {journal} {Phys. Rev. A}\ }\textbf {\bibinfo {volume} {92}},\
		\bibinfo {pages} {063841} (\bibinfo {year} {2015})}\BibitemShut {NoStop}%
	\bibitem [{sup()}]{supplement}%
	\BibitemOpen
	\href@noop {} {}\bibinfo {note} {See Supplemental Material at [url] for
		additional details of the calculations.}\BibitemShut {Stop}%
\end{thebibliography}
\end{document}